\documentclass[twocolumn,showpacs,preprintnumbers,amsmath,amssymb]{revtex4}

\usepackage{graphicx}
\usepackage{dcolumn}
\usepackage{bm}
\begin{document}
                                                                                                                             
\preprint{CUPhysics/02/2009}
\title
{Persistence of unvisited sites in presence of a quantum random
walker}

\author{Sanchari Goswami and Parongama Sen}
\affiliation{
Department of Physics, University of Calcutta,92 Acharya Prafulla Chandra Road,
Calcutta 700009, India.\\
}

\begin{abstract}
A study of persistence dynamics is made for the first time in a quantum system
by considering the dynamics of a quantum random walk.
For a discrete  walk on a line starting at $x=0$ at time $t=0$, 
the persistence probability $P(x,t)$ that  a  site at $x$ has not been visited 
till time $t$ has been calculated.
 $P(x,t)$ behaves as  $(t/|x|-1)^{-\alpha}$ with $\alpha \sim 0.3$ while    
the global fraction ${\cal{P}}(t) = \sum _xP(x,t)/2t$ of sites remaining unvisited at time $t$ 
attains a constant value.
$F(x,t)$,   
 the probability that  the site at $x$ is visited for the first time at $t$
behaves as $(t/|x|-1)^{-\beta}/|x|$ where $\beta = 1+ \alpha$ for  $t/|x|>> 1$,
 and ${\cal{F}}(t) =\sum _xF(x,t)/2t \sim 1/t$.
A few other properties related to the persistence and 
first passage times are studied and  some fundamental  differences  
between the classical and the quantum 
cases are observed. 
\end{abstract}

\maketitle
                                                                                                                            
Various studies on persistence in dynamical systems have 
been made in 
recent years \cite{satya}  where the dynamical process is purely classical in nature. The persistence
probability that   the order parameter 
in a magnetic system  has not changed 
sign till the present time \cite{derrida} and the persistence of unvisited sites
in a diffusion problem  \cite{bray} are common examples which have received   
extensive interest.
The importance of persistence phenomena lies in the fact that the persistence
probability  
in many systems
 shows an algebraic decay with a novel exponent.

The dynamics of a quantum system is expected to be different 
from the corresponding classical case. However, to the best of our knowledge 
no attempt to study persistence in quantum systems have been made so far.
Here we report a   study of persistence and related dynamical 
features in a quantum system
which has a classial analogue such that it is possible to 
determine the differences, if any, between them.  

The dynamical process we consider is a discrete quantum random walk (QRW).
Classical random walk (CRW) on a line is a much studied topic \cite{chandra,book,redner}
where at every step
one tosses a fair coin and takes a step, either to the left or right. 
In a quantum generalisation  of a 
classical walk, one may consider superposition
of particle movement in both directions, however such a process is physically
impossible due to non-unitarity.  
 A  QRW in one dimension thus involves a quantum particle
on a line with an additional degree of freedom which may be called chirality \cite{nayak,kempe}.
The chirality  takes values left and right analogous to Ising spin states $\pm 1$, and directs the motion of the particle. 
At any given time, the particle is in a superposition of left and right
chirality. Thus the wave function has two components and at each time step,
the chirality undergoes a rotation (a unitary transformation in general) 
and the particle moves according to its final chirality state. 

The reason to study the qunatum random walk is that it is already known
 to have some distinctive behaviour when compared to the classical walk.
While in  the CRW,  
after $t$ time steps, the variance $\sigma^2 \propto t$, the quantum walker 
propoagates  much faster with 
$\sigma \propto t$.  
This feature of faster flow of a quantum walker has made its application
in quantum computation highly relevant.
Moreover, the QRW spreads 
roughly uniformly over the interval
$[-t/\sqrt{2}, t/\sqrt{2}]$ whereas in the 
classical case  the distribution is peaked at the origin and drops off 
exponentially fast.

In the case of the  classical random walker, the well studied 
quantities are   (a) the first passage time $F_{cl}(x,t)$ which is the probability
that the walker has reached the site $x$ for the first time at time $t$ \cite{redner} and (b) 
persistence   or the
survival probability $P_{cl}(x,t)$  defined as  
the probability that the  site at $x$ has not been visited till time $t$ \cite{chandra,book}. 
A related quantity is the average persistence probability
 given by $\sum_x P_{cl}(x,t)/L$ (where $x$ can take values from $-L/2 $ to
$L/2$). This quantity is global and can be interpreted as the fraction of  
 sites remaining unvisited till time $t$. 

%

In this paper, we have studied both the first passage probabilities and
the persistence probabilities of sites in presence of a quantum random walker.
Not only do 
we  find out    $F(x,t)$ and $P(x,t)$ as functions of $t$ and $x$, we also
observe  several other interesting features of the distributions which
indicates that there are two independent dynamical
exponents for the quantum random walker. 

In this context, it may be mentioned that in the QRW, the walks 
can be realized in several possible ways, e.g., a walk can be infinite timed in which  the walk is allowed upto time $t$ at which all 
measurements are made. Another method may be to have a semi infinite 
walk where the question is asked at every time step whether the   
walker is in a specific location, 
and if it is so, the system is allowed to collapse to that state. In our case, 
we simply evaluate the occupation probabilities  
 at each step for each
discrete location and calculate the persistence and first passage probabilities
which are related to it. There  is no actual measurements being made 
which will lead to a collapse to any particular state. Rather,
our calculations will correspond to what would be the result of 
repeated experiments. 


The states for a quantum random walker are written as  $|x\rangle |d\rangle$
where $x$ is the location   in real space             
and
the $d$ the chirality having  either  ``left'' and ``right'' values. The 
chiral states are  denoted as $|L\rangle$ and $|R\rangle$. 
A conventional
choice of the unitary operator causing 
the rotation of the chirality state is the  Hadamard coin represented by
\begin{equation}
 H=\frac{1}{\sqrt{2}}\begin{pmatrix}
    1 & 1\\
    1 & -1
   \end{pmatrix}. 
\end{equation}

The rotation is followed by a  translation represented by the operator $T$:
\begin{equation}
\begin{array}{l}
T |x\rangle |L\rangle \rightarrow   |x-1\rangle|L \rangle\\
T |x\rangle |R\rangle \rightarrow   |x+1\rangle|R \rangle\\
\end{array}.
\end{equation}

The two component  wave function $\psi(x,t)$ describing the position of the particle is written as 
\begin{equation}
\psi(x,t)=
  \begin{pmatrix}
    \psi_{L}(x,t)\\
    \psi_{R}(x,t) 
  \end{pmatrix},
\end{equation}
and the occupation probability is given by
\begin{equation}
f(x,t) = |\psi_{L}(x,t)|^2 + |\psi_{R}(x,t)|^2.
\end{equation}

Normalisation of the probability implies that at every step $\sum_x f(x,t) = 1$.

\smallskip
We have used two methods to evaluate $f(x,t)$: \\\
1. A quantum random walk is generated using the above operators and 
 $\psi_{L}(x,t)$ and  $\psi_{R}(x,t)$ are evaluated numerically 
for all $x$ and $t$. 
 We start with the 
walker at the origin with 
$\psi_{L}(0,0) = a_0,  \psi_{R}(0,0)= b_0; ~~a_0^2 + b_0^2 =1$ 
and all other $\psi_{L}$  and  $\psi_{R}$ taken equal to zero.\\ 
\smallskip
2. In the second method we use the expressions \cite{nayak}

\begin{equation}
 \psi_{L}(x,t)=\frac{1+(-1)^{x+t}}{2}\int\frac{dk}{2\pi}(1+\frac{\cos k}{\sqrt{1+\cos^{2}k}})e^{-i(\omega_{k}t+kx)}
\label{analytic1}
\end{equation}

\begin{equation}
 \psi_{R}(x,t)=\frac{1+(-1)^{x+t}}{2}\int\frac{dk}{2\pi}\frac{e^{ik}}{\sqrt{1+\cos^{2}k}}e^{-i(\omega_{k}t+kx)}
\label{analytic2}
\end{equation}
(which are obtained for a initial state with left chirality, i.e., $a_0=1, b_0=0$)
and  evaluate $f(x,t)$ directly.
\begin{figure}
\noindent \includegraphics[clip,width= 4cm, angle=270]{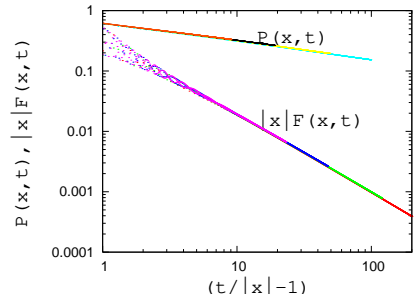}
\caption{The data collapse for the persistence probability and scaled 
first passage probability for a quantum random walk shown for different values of $x$.}
\label{collapse1}
\end{figure}

\begin{figure}
\noindent \includegraphics[clip,width= 6cm]{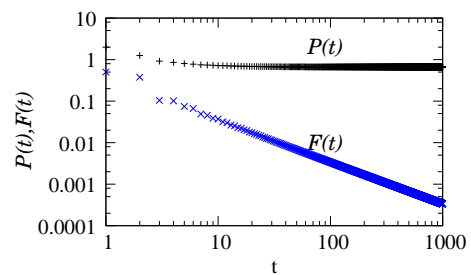}
\caption{Plot of average  persistence ${\cal{P}}(t)$ and first passage time ${\cal{F}}(t)$ indicate  ${\cal{P}}(t)$ is approaching a constant for large time whereas ${\cal{F}}(t)$ behaves as   $t^{-1}$.
}
\label{tot_pers_fp}
\end{figure}


The persistence probability that a site $x$ has not been visited till time $t$, is  given by in terms of the occupation probabilities:
\begin{equation}
P(x,t)=\prod_{t'=1}^{t}(1-f(x,t^{'})),
\end{equation}
while the first passage time
\begin{equation}
F(x,t)=\prod_{t'=1}^{t-1}(1-f(x,t^{'}))f(x,t)
\end{equation}
$$=P(x,t-1)f(x,t).$$

Some time dependent features like hitting time, recurrence time etc. \cite{hitting,Stefanak} of the QRW have been studied earlier which involve the first passage time. However, in these studies, the spatial dependence has not been considered. For example, quantities like first passage time specifically at the origin has been estimated.

In the first method of calculation,  in general, real values of the  coefficients  $a_0$ and $b_0$ give an asymmetric result 
for $f(x,t)$ in the sense  the occupations probabilities $f(x,t) \neq f(-x,t)$ 
as there is destructive interference. 
However, taking  specific complex values of $a_0$ and $b_0$ or by taking a different coin results in a symmetric walk \cite{kempe}.

We first discuss the results for the symmetric case obtained by taking  $a_0=1/\sqrt{2}, b_0=i/\sqrt{2} $.
We find that  $P(x,t)$  and $F(x,t)$ as  functions of  $t$ and $x$ behave as 
\begin{equation}
P(x,t) \propto (t/|x|-1)^{-\alpha}
\label{perseq}
\end{equation}
for $t/|x| > 1$,  
and 
\begin{equation}
F(x,t) \propto (t/|x|-1)^{-\beta} /|x|
\label{firsteq}
\end{equation}
for  $t/|x| >> 1$  with 
$\alpha \simeq 0.31$ and $\beta \simeq 1.31$. 
The data collapse for $P(x,t)$ and $F(x,t)$ are shown in Fig. \ref{collapse1}.

Using the above form of $P(x,t)$, the average probability ${\cal{P}}(t)$ that a site is unvisited
till time $t$ is given by
$\sum_{x=-t}^{t} P(x,t)/2t$ (since  $x$ can assume  discrete values from $-t$ to $+t$)
 which in the symmetric case is approximately equal to
\begin{equation}
\frac{1}{t}\int_0^t P(x,t) dx = \frac{A}{t}  \int_0^t (\frac{t}{x} -1)^{-\alpha}dx.
\label{avper}
\end{equation}
The integral on the rhs of eq (\ref{avper}) can be done exactly giving a
constant value equal to 0.738
 (using $A=0.626$ which we get from the fitting of $P(x,t)$ 
shown in Fig.\ref{collapse1}),
 indicating that  ${\cal{P}}(t)$ is time independent.
We indeed find by numerically evaluating $\sum_{x=0}^{t} P(x,t)/t$ that it
approaches a constant value as $t$ becomes large. The constant is  $\approx 0.66$,
quite close  to the value obtained above. On the other hand,
$\sum_{x=-t}^{t} F(x,t)/2t$ varies as $t^{-1}$, a result which can be obtained using a similar approximation 
and
also obtained numercially. 
The numerical results are shown in Fig. \ref{tot_pers_fp}.

\begin{figure}
\noindent \includegraphics[clip,width= 4.2cm, angle=270]{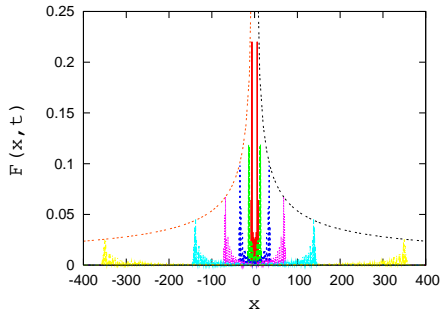}
\caption {Typical variation of $F(x,t)$ against $x$ for 
different values of $t$. The peaks are approximately fitted to $0.81x_{max}
^{-0.59}$.}
\label{fmax_x}
\end{figure}

\begin{figure}
\noindent \includegraphics[clip,width= 4.2cm, angle=270]{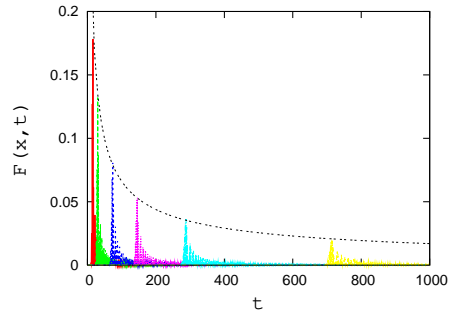}
\caption{Typical variation of $F(x,t)$ against $t$ for different values of $x$.  The peaks are approximately fitted to $t_{max}^{-0.59}$.}
\label{fmax_t}
\end{figure}

The function  $F(x,t)$ in fact has some additional features. Plotting $F(x,t)$ against $x$ or 
$t$, we notice that it has an oscillatory behaviour. These oscillations which die down for large values
of $t/x$ as is apparant from Fig. \ref{collapse1} can be traced to 
the oscillatory behaviour of $f(x,t)$ for a QRW observed earlier  \cite{nayak,kempe}.    
 From Figs \ref{fmax_x} and \ref{fmax_t}, we observe that $F(x,t)$ actually attains a 
maximum value $F_{max}(x,t)$ at values of $|x|=x_{max}$ (or $t=t_{max}$) for fixed values of $t$ (or $x$).
We notice that 
$F_{max}(x_{max},t) \propto x_{max}^{-\delta}$ where $\delta \simeq 0.59$.
Keeping $x$ fixed,  $F_{max}(x,t_{max})$ versus
$t_{max}$ shows the  same kind of dependence, i.e., $F_{max}(x,t_{max}) \propto t_{max}^{-\delta}$. That the scalings with $t_{max}$ and $x_{max}$ turn out to be identical  is 
not surprising as $x$ scales as  $t$ in a QRW.
 It is not possible to obtain this scaling form directly from eq. (\ref{firsteq})   since
 $F(x,t)$ attains a
maximum value when  $t/|x|$ is close to unity where the   fitted scaling form is not exactly  valid.
In fact, eq. 
(\ref{firsteq})  does not give any maximum value at all.


Another dynamic quantity called  hitting time has been estimated earlier for the QRW,
in which an  absorber is assumed to be located at a  specific vertex of a hypercube within which the
walk is conceived \cite{hitting}. The average hitting time is by definition the average time to reach that
particular vertex for the first time. 
One can  evaluate the
average hitting time $\tau_h(x)$ using  $\tau_h(x) = \sum_0^T t F(x,t)$
where $t$ is allowed to vary from $0$ to $T$:
\[
\tau_h \approx \int_0^{T}t F(x,t)dt
\]
\[
\sim {T}^{2-\beta} x^{\beta -1}/(2-\beta) + O({T}^{-{\beta+1}}).
\]
The numerical data (not shown) gives a fairly good agreement with this scaling. 
The above equation shows that  $\tau_h$ blows up for $T \to \infty$ in agreement with some earlier results using other 
coins \cite{hitting}. 

When $a_0$ and $b_0$ are allowed to take up real values, in general the
probability $f(x,t)$ is asymmetric.
Now we study the  behaviour of $P(x,t)$ and $F(x,t)$ for $x>0$ and $x < 0$
independently.
In particular, we find that when there is a left bias, i.e., the probability
that $f(x<0,t) > f(x>0,t)$,
$P(x<0,t)$  is less than $P(x>0,t)$, which is to be expected. Moreover,
the scaling behaviour for $x < 0$ shows an additional dependence on $x$,
$P(x<0,t)$ is now given by
\[
P(x,t) \propto (t/|x|-1)^{-\alpha}|x|^{p}
\]
where $p$ is approximately 0.03 for $a_0=1$, $b_0=0$.
If $a$ and $b$ are chosen differently, the value of $p$
 shows up a variation with the chosen values, however it is still $O(0.01)$.
Such a small value of $p$ suggests that this could be due to
numerical errors. Fig. 5 shows the left-right persistence behaviour.

\begin{figure}
\noindent \includegraphics[clip,width= 4.5cm, angle=270]{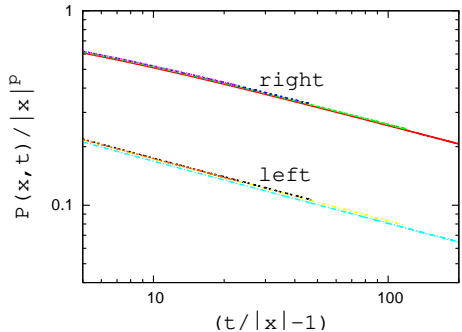}
\caption{Data collapse of $P(x,t)$ for a quantum random walk with 
a left bias. For  the right side data,  $p=0$ and for the left side,  $p=0.03.$}
\end{figure}

\begin{figure}
\noindent \includegraphics[clip,width= 4.5cm, angle=270]{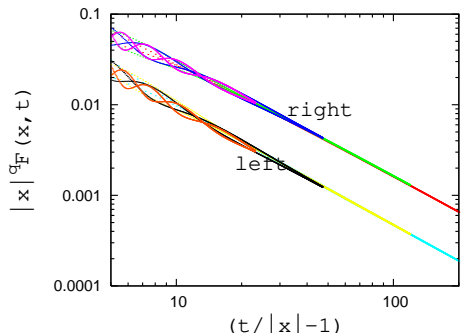}
\caption{Data collapse of $F(x,t)$ for a quantum random walk with 
a left bias. For  the right side data,  $q=1$ and for the left side,  $q=1-0.07$}
\end{figure}

The first passage probability for $x < 0$ shows a similar correction to scaling
\[
F(x,t) \propto (t/|x|-1)^{-\beta}|x|^{-q}
\]
where  $q = 1-  O(0.01)$
Fig. 6 shows typical variations of $F(x,t)$ for $a_0=1$, $b_0=0$ with $q=0.93$ .

The significance of the various results obtained in the present work becomes quite clear when these are compared
to those of the classical random walker.
For a CRW, the first passage time $F_{cl}(x,t)$ is known exactly \cite{chandra,book}
\[
F_{cl}(x,t) = {\rm{const}}\frac{|x|}{t^{3/2}} \exp(-x^2/Dt),
\]
where $D$ is a diffusion constant.
Hence $F(x,t)$ behaves as $t^{-3/2}$ for large $t$.
The persistence or survival probability is given by 
${\rm{erf}}(\frac{|x|}{\sqrt{Dt}})$ which behaves  as $|x|t^{-1/2}$
as $t \to \infty$. 
In fact the
two quantities, persistence probability and first passage time are related in the continuum limit:
\begin{equation}
F(x,t)=-\frac{\partial P(x,t)}{\partial t},
\label{relation}
\end{equation} 
and hence the scaling behaviour (with time) of   $F$  can be found out  from that of  $P$ or vice versa;
e.g., $\beta_{cl} = \alpha_{cl} + 1$, 
where $P_{cl} \propto t^{-\alpha_{cl}}$ 
and $F_{cl} \propto t^{-\beta_{cl}}$.
 
In general one can write $x \propto t^\gamma$ for these kind of dynamical
phenomena where the inverse of $\gamma$ is a dynamical exponent.  It can 
be expected that the  scaling behaviour of all other quantities will be 
dictated by the exponent $\gamma$. 
Here we would like to emphasise that while this is true for the classical case,
for the QRW, that is not the case. Let us consider each physical measure
for the CRW and QRW to establish this claim.

In  the CRW, $\gamma = \gamma_{cl}=1/2$ and the scaling of the first passage 
time with time occurs with an exponent $ \beta_{cl} = 3\gamma_{cl}$ and 
the persistence probability with an exponent $\alpha_{cl} = \gamma_{cl}$.
For the quantum walker, $\gamma = \gamma_q =1$,  
 however, 
the persistence behaviour
and the first passage times vary in time with exponents $\alpha$ and $\beta$ which are not simple
multiples of $\gamma_q$  although  $F(x,t)$ and $P(x,t)$ indeed obey the relation given in eq. (\ref{relation})
such that $\beta = 1+ \alpha$.
The form of the persistence probability and first passage time are in fact entirely different from those of the classical case
as indicated by the collapsed data.

 In the classical case, the global fraction ${\cal{F}}_{cl}$  
behaves as $t^{-1}$. From this, one can infer that 
the global fraction ${\cal{P}}_{cl}$   is 
approximately a constant. In the quantum case, surprisingly, we have 
identical scaling behaviour for ${\cal{P}}(t)$ (which saturates to
a constant value) and ${\cal{F}}(t)$ (which varies as $t^{-1}$). In terms of 
$\gamma$ however, ${\cal{F}}_{cl} \propto  t^{-2\gamma_{cl}}$ while  
${\cal{F}} \propto t^{-\gamma_q}$.   
In the QRW  a striking feature is that these global properties 
do not involve the exponents $\alpha$ and $\beta$ at all 
 and only 
$\gamma_q$  dictates their behaviour. 

Concerning the     maximum values of the classical probability $F_{cl}$,
we find that it behaves as $1/t_{max}$ (for $x$ constant) or $1/x_{max}^2$ (for 
$t$ constant) showing that 
the obtained exponents are  
simple multiples of $\gamma_{cl} =1/2$. 
On the other hand,
the behaviour of $F_{max}$ in the quantum case appears to 
 depend on the 
value of $\alpha$  and not $\gamma_q$ as it varies with  ${t_{max}}$ or 
$x_{max}$ with an exponent  $\delta$ which is very close to 
$2\alpha$ numerically. 

The average hitting time for
a CRW  is found to vary as $T^{\gamma_{cl}}$. 
In the QRW, this variation is given by $T^{2-\beta}$. For the classical case,
$2-\beta_{cl} = \gamma_{cl}$  but since no such relation exists for the
quantum case, the hitting time scaling is therefore {\it{not}} dictated by
$\gamma_q$ but by  $\beta$ (or $\alpha$) only.

Thus the scaling 
forms of different quantities for the quantum and classical
walks are not identical in general and
the scaling behaviour  in the quantum case is dictated either by $\gamma_q$ or
by $\alpha$.
From this we draw the important  conclusion   that there are two nontrivial and independent
exponents for the QRW in contrast to the CRW.
This result may have serious impact on
quantum computational aspects.


Acknowledgement: Financial supports from DST grant no. SR-S2/CMP-56/2007 (PS)  and 
UGC sanction no. UGC/209/JRF(RFSMS) (SG) are acknowledged.

\end{document}